\documentclass[aps,preprint,pra,floatfix]{revtex4}
\usepackage{graphicx}

\begin{document}
\title{Modelling magnetism of C at O and B monovacancies in graphene}
\author{T.\ P.\ Kaloni$^1$, M.\ Upadhyay Kahaly$^1$, R.\ Faccio$^{2,3}$, and U.\ Schwingenschl\"ogl$^{1,}$}
\email{Corresponding author. Tel.: +966 544700080. E-mail address: udo.schwingenschlogl@kaust.edu.sa (U.\ Schwingenschlogl)}
\affiliation{$^1$PSE Division, KAUST, Thuwal 23955-6900, Kingdom of Saudi Arabia\\
$^2$Crystallography, Solid State and Materials Laboratory (Cryssmat-Lab), DETEMA, 
Facultad de Qu\'imica, Universidad de la Rep\'ublica, Gral. Flores 2124, P.O. Box 1157, 
Montevideo, Uruguay\\ 
$^3$Centro NanoMat, Polo Tecnol\'ogico de Pando, Facultad de Qu\'imica, Universidad de 
la Rep\'ublica, Cno. Aparicio Saravia s/n, 91000, Pando, Canelones, Uruguay}

\begin{abstract}
The presence of defects can introduce important changes in the electronic
structure of graphene, leading to phenomena such as C magnetism. In addition, vacancies
are reactive and permit the incorporation of dopants. This paper discusses the electronic
properties of defective graphene for O and B decoration. Phonon calculations
allow us to address directly the stability of the systems under study. We show
that it is possible to obtain magnetic solutions with and without dangling bonds,
demonstrating that C magnetism can be achieved in the presence of B and O.
\end{abstract}

\maketitle

{\bf 1. Introduction}

C nanostructures have attracted the attention of the scientific community because of 
both unique fundamental properties and potential in technological applications. 
In particular, the high coherence length and large conductivity \cite{castro} of 
C materials are promising in electronics and spintronics. On the other hand, also
magnetism has been observed in this class of materials \cite{Palacio}, generating huge
interest of both experiment and theory \cite{han,Setze,rode,lehtinen,Flipse1,Ugeda,mousumi}.
The structural, electronic, and magnetic properties of defective graphene have been
studied in detail and it has been confirmed that vacancies are the source of magnetism
\cite{new1,new2,new3,new4}. It has been demonstrated that point defects lead to notable
paramagnetism but no magnetic ordering is achieved down to liquid helium temperature
\cite{nair}. In addition, the point defects carry magnetic moments irrespective of the vacancy
concentration. In \cite{palacio1} the authors report that the extended $\pi$-band magnetism
reduces to zero in the limit of monovacancies in graphene. For these reasons, it is
important to evaluate the role of saturation and chemical modifications in the
neighborhood of mono and multivacancies in order to understand how a persistent
magnetic ordering can be achieved.

Graphene samples in general are characterized by the presence of $sp^{2}$ hybridized C atoms, 
extended pores, and various attached functional groups, as revealed by scanning electron
microscopy, atomic force microscopy, and Raman spectroscopy. It has been concluded that the
appearance of magnetism requires the presence of defects, adatoms, or topological defects
\cite{khana}. Zagaynova and coworkers have prepared magnetic 
C by chemical oxidation \cite{mombr}. B doping in general leads to different
magnetic responses depending on the amount of doping. This behavior has been partially addressed by 
Faccio and coworkers \cite{pardo2}, who have demonstrated that magnetism
requires the proximity of B and a monovacancy. However, when the B atom is too close to
the vacancy all the dangling bonds are reconstructed and magnetism is suppressed. 
The authors of \cite{kaloni} have addressed the situation of multiple O atoms
attached to a monovacancy in graphene, demonstrating that vacancies are magnetic if and
only if they are metallic and non-magnetic if and only if they are semiconducting.
Metallicity and magnetism thus are simultaneously determined by the presence or absence
of dangling C bonds after the oxidation. In our present work, we give a 
theoretical study of the effects of vacancies interacting with different amounts of B and O 
atoms, focusing on the structural, electronic, and magnetic properties. An experimental
realization of the proposed structures in possible along the lines of Ref.\ \cite{carbon}.

{\bf 2. Methods}

We employ density functional theory in the generalized gradient approximation
(Perdew-Burke-Ernzerhof scheme) as implemented in the Quantum-ESPRESSO code \cite{paolo}.
All calculations are performed with a plane wave cutoff energy of 544 eV. A Monkhorst-Pack
$8\times8\times1$ k-mesh is used to relax the structures and a $16\times16\times1$ k-mesh
to calculate the density of states (DOS) with a high accuracy. We use a $5\times5$ supercell
of pristine graphene in our calculations \cite{cheng1}. This supercell size has shown in
various studies to be sufficient for functionalization of graphene by atoms, small
molecules, amine and nitrobezene groups, and others \cite{new12,new13,new14,new15,new16,new17}.
Our supercell contains 50 C atoms and has a lattice constant of $a = 12.2$ \AA\ with a 
20 \AA\ thick vacuum layer on top. The atomic positions are relaxed upto an energy
convergence of $10^{-7}$ eV and a force convergence of 0.005 eV/\AA. For our vibrational
study, the phonon frequencies and eigenvectors at the $\Gamma$-point are calculated
with an energy error of less than $10^{-14}$ eV. Phonon frequencies at the $\Gamma$-point
are determined by density functional perturbation theory for evaluating the structural
stability \cite{Mod.}. We note that the numerics break the
symmetry of the dynamical matrix and introduce slight errors in the phonon frequencies
of maximal 15 cm$^{-1}$, which, however, are not critical.

\begin{figure*}[t]
\includegraphics[width=0.77\columnwidth]{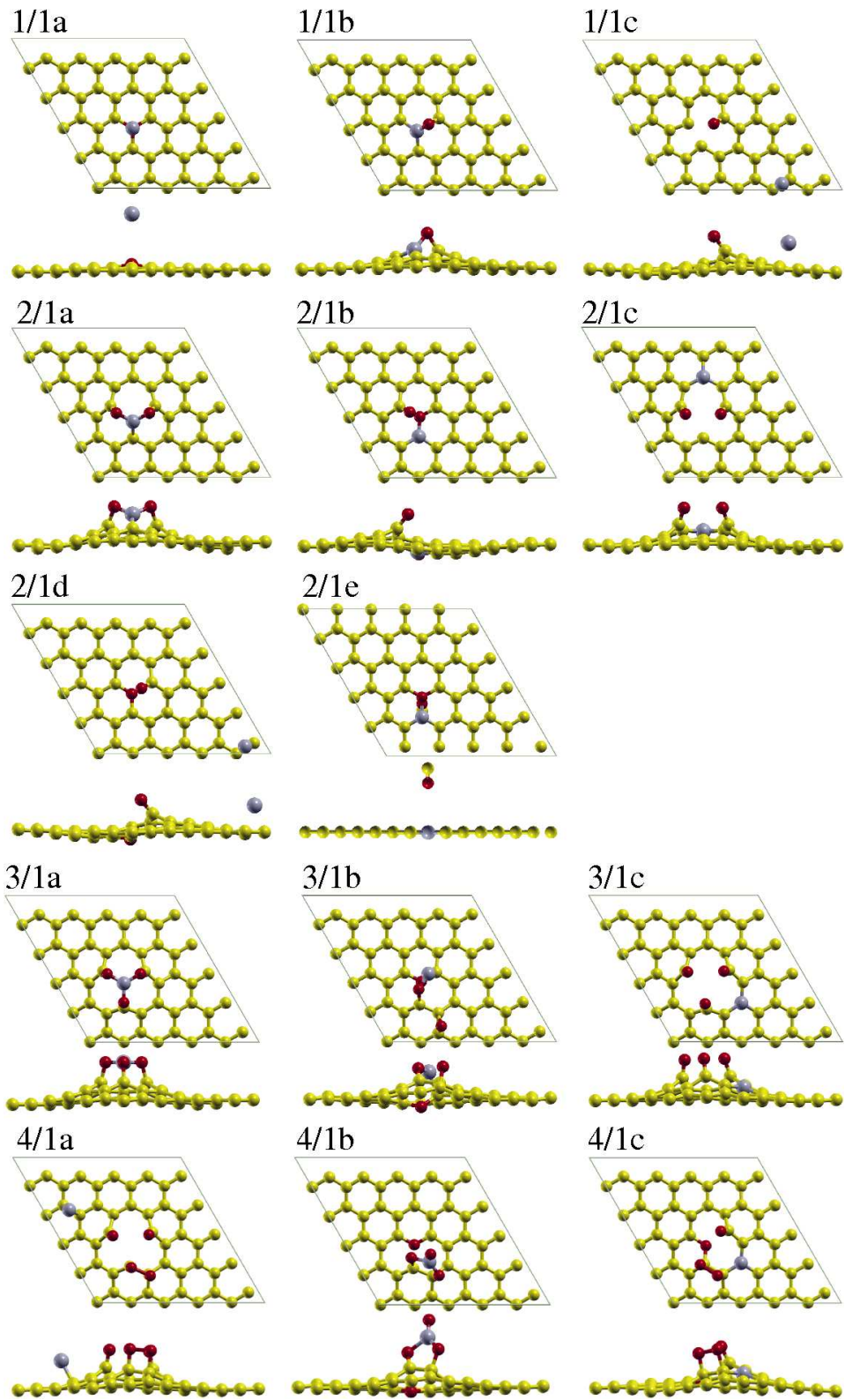}
\caption{\label{fig1} Crystal structures for 1 B and 2, 3, or 4 O atoms adsorbed at a
monovacancy in graphene. The yellow, red, and gray spheres represent C, O, and B, respectively.}
\end{figure*}

\begin{figure*}[t]
\includegraphics[width=0.5\columnwidth]{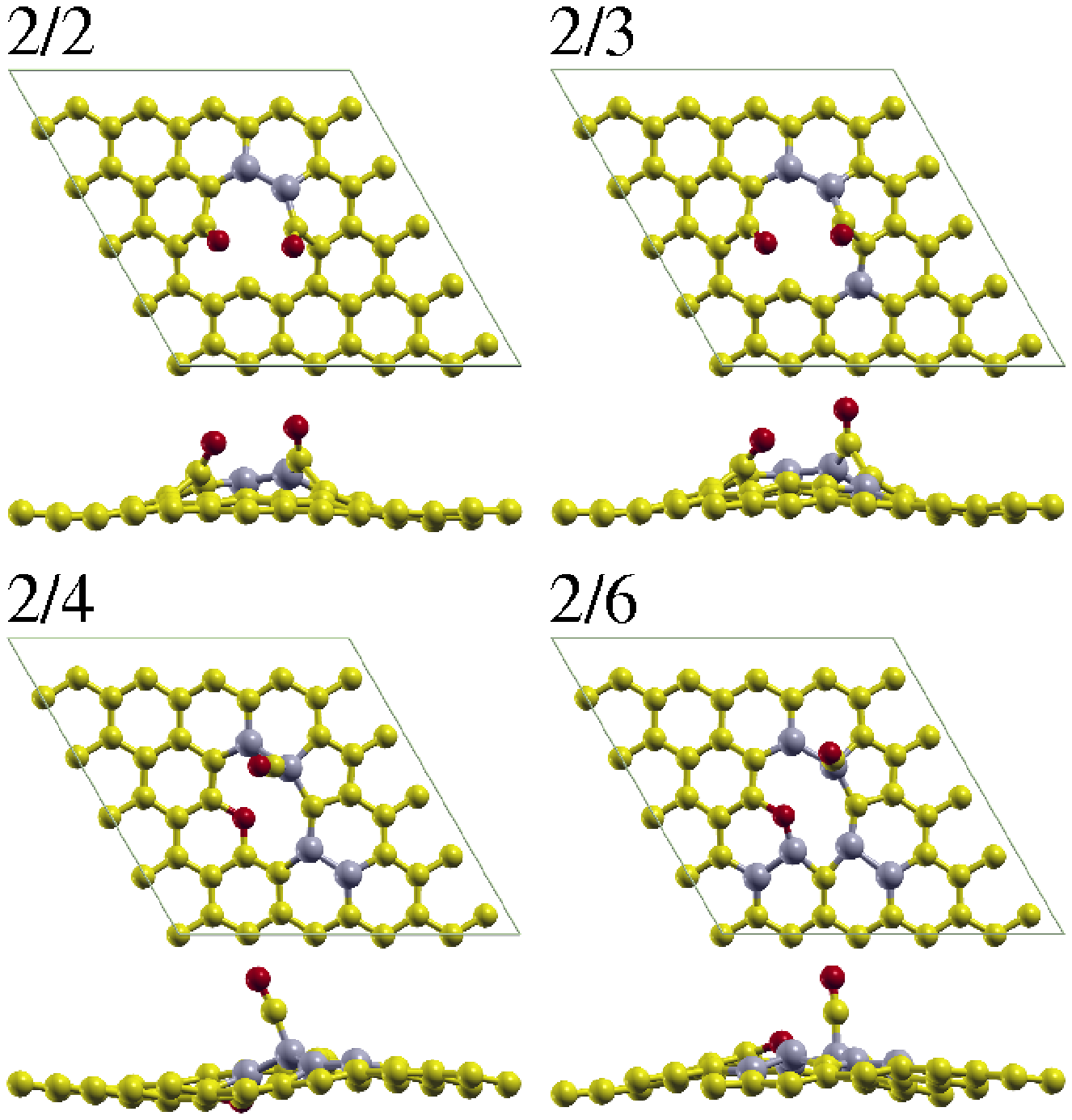}
\caption{\label{fig2} Crystal structures for 2 O and 2, 3, 4, or 6 B atoms adsorbed at a
monovacancy in graphene. The yellow, red, and gray spheres represent C, O, and B, respectively.}
\end{figure*}

In a first step, we create a monovacancy in our $5\times5$ supercell and relax the system. 
Afterwards, we add O and B atoms in the vicinity of the monovacancy for different
starting geometries and relax the system again. Several prototypical monovacancies are
prepared in order to consider various possibilities with different O and B concentrations. 
We name these configurations giving first the number of O atoms and second
the number of B atoms. For example, configuration 1/2 has 1 O and 2 B atoms in the
vicinity of the monovacancy. All the configurations under study are shown in Figs.\ 1 and 2.
Note that we have tested the required supercell size by addressing for the most stable
configurations $10\times10$ supercells and find no significant modification in the induced
buckling of the graphene sheet after O and B adsorption.

{\bf 3. Results and discussion}

At a clean monovacancy the under-coordinated C atom is subject to Jahn-Teller distortion
\cite{Jahn} and therefore moves out of the graphene plane by 0.13 \AA. We find a total
magnetic moment of 1.35 $\mu_B$, where 1 $\mu_B$ is due to the localized dangling bond
of one C atom and the remaining moment is carried by extended $\pi$ states \cite{kaloni}. 
The vacancy formation energy, $E_{form}$, is determined by the expression 
\begin{equation}
E_{form}=E_{vacancy}-\frac{N-1}{N}E_{graphene},
\end{equation}
where $E_{vacancy}$ and $E_{graphene}$ are the total energies of defective and pristine
graphene, respectively. Moreover, $N$ is the number of atoms in the pristine graphene.
The obtained value of $E_{form}$ is 7.4 eV, which agrees well with previous reports
\cite{pandey,eggie}. 

We first study a monovacancy decorated by 1 O atom to which 1 B atom is added. 
When forcing the O atom to occupy a position within the graphene plane connected
to 3 C atoms the B atom is released and repelled, leading to configuration 
1/1a. In configuration 1/1b the B atom connects to 1 O atom and 2 C atoms, saturating all
dangling bonds. Moreover, configuration 1/1c deals with B adsorption far from the
vacancy, for which we obtain a bridge position over a C$-$C bond with $d_{B-C}=1.87$ \AA.
The energetics of the doping process indicate that the chemical reaction between
graphene-oxide and B can be evaluated as
\begin{equation}
E_{form}=E_{vacancy+B+O}-\frac{N-N_C}{N}E_{graphene}-\frac{N_O}{2}E_{O_2}-N_BE_{B,bulk},
\end{equation}
where $E_{vacancy+B+O}$ is the total energy after B adsorption on graphene-oxide.
In addition, $N_C$, $N_O$, and $N_B$ are the numbers of missing C atoms and additional O and B
atoms in the decorated system. It turns out that configuration 1/1b is energetically favorable
with $E_{form}=-6.9$ eV, see Table I, while for configurations 1/1a and 1/1c we obtain in
each case a value of $-1.7$ eV. By the huge energy
difference it is clear that only configuration 1/1b will be realized. Due to the B
adsorption and the induced C and O reconstructions, this configuration becomes a metal,
see Fig.\ 3(a). The Dirac cone, which is still due to the C $p_z$ orbitals with no
contribution of O or B, splits and shifts above $E_F$ by about 0.6 eV. Interestingly,
a magnetic moment of 0.2 $\mu_B$ is observed although no dangling bonds are present.

\begin{table*}[t]
\begin{tabular}{ccc} 
$\quad$System$\quad$  & $\quad$$E_{form}$ (eV)$\quad$ &$\quad$Magnetic moment ($\mu_B$)$\quad$  \tabularnewline      
\hline
1/1a    &$-$1.7              &1.0                \tabularnewline                                      
1/1b    &$-$6.9            &0.2               \tabularnewline              
1/1c    &$-$1.7              &0.0              \tabularnewline  
\hline
2/1a    &$-$7.0             &0.0             \tabularnewline                                       
2/1b    &$-$7.4             &0.1            \tabularnewline              
2/1c    &$-$5.3             &1.0            \tabularnewline
2/1d    &$-$4.8             &0.0            \tabularnewline              
2/1e    &$-$6.9             &0.5$^*$        \tabularnewline 
\hline
3/1a    &$-$4.4             &0.2           \tabularnewline                                       
3/1b    &$-$3.4             &1.0          \tabularnewline              
3/1c    &$-$3.5             &1.0          \tabularnewline
\hline 
4/1a    &1.1             &0.0          \tabularnewline                                        
4/1b    &$-$11.3             &0.0          \tabularnewline              
4/1c    &1.6             &0.0           \tabularnewline
\hline
2/2   &$-$8.4              &1.0            \tabularnewline                                       
2/3    &$-$12.8             &1.0            \tabularnewline              
2/4    &$-$22.9             &0.0           \tabularnewline
2/6    &$-$34.4             &0.0            \tabularnewline
\end{tabular}\caption{Formation energy and magnetic moment. 
$^*$The magnetic moment is located on the released CO molecule.} 
\end{table*}

\begin{figure*}[t]
\includegraphics[width=1\columnwidth]{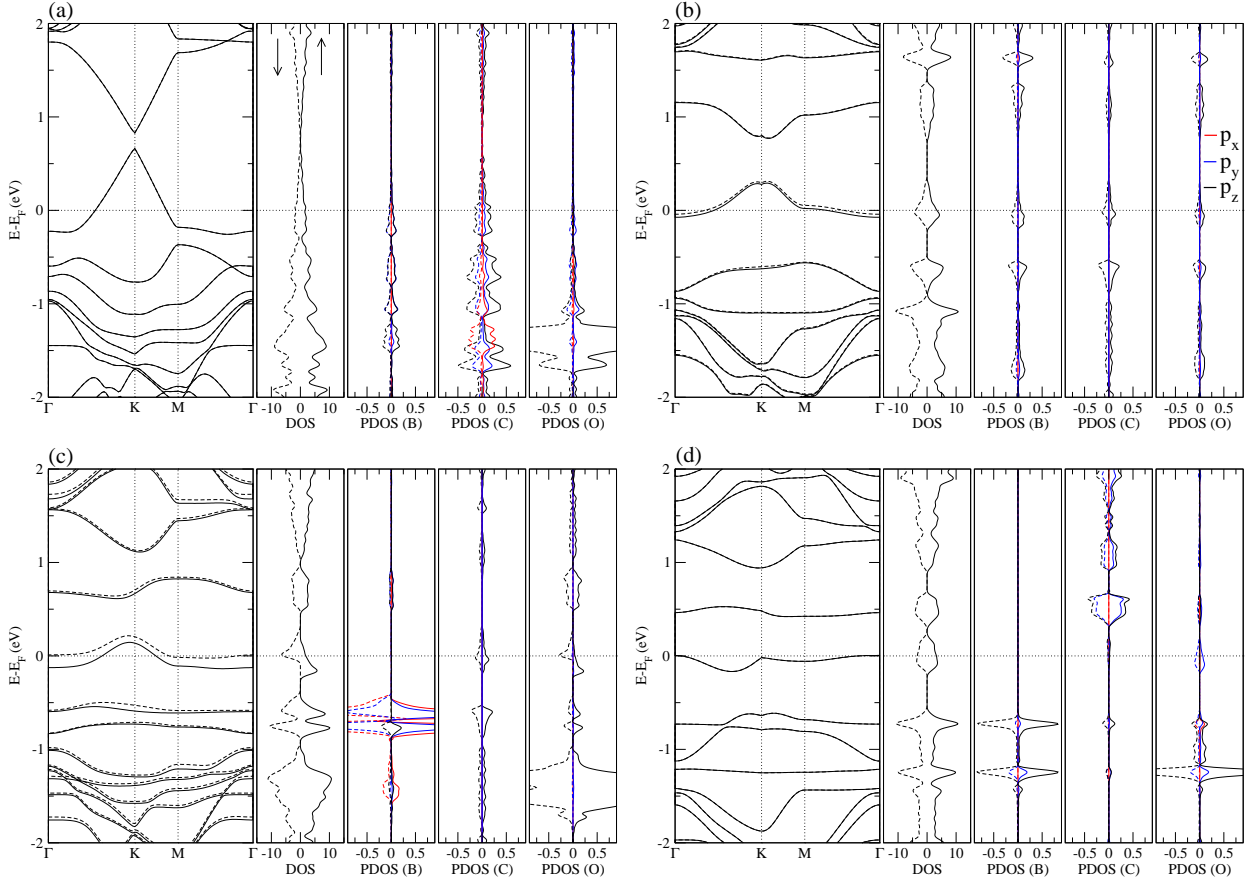}
\caption{\label{fig3} Electronic band structure and DOS for the structures
(a) 1/1b, (b) 2/1b, (c) 3/1a, and (d) 4/1b.}
\end{figure*}

Turning to an oxidized monovacancy decorated by 2 O atoms, in
configuration 2/1a the B atom establishes strong interaction with O and C, which originally
was under-coordinated. We obtain $E_{form}=-7.0$ eV and no spin polarization due to
saturation of all dangling C bonds. In configurations 2/1b and 2/1c the B atom is adsorbed
at the boundary of the vacancy, leading to reconstructions with one and two C=O bonds,
respectively. We obtain for $E_{form}$ values of $-7.4$ eV and $-5.3$ eV, where
configuration 2/1c is less favorable due to the presence of a dangling C bond which
induces a magnetic moment of 1 $\mu_B$.
In configuration 2/1d the B atom is adsorbed far away from the monovacancy, which yields
a surprising geometrical reconstruction: the two C=O bonds change into one ketone (C=O)
and one ether group. However, with $E_{form}=-4.8$ eV this configuration is not favorable.
Finally, in configuration 2/1e the B atom is located in the graphene plane and a CO
molecule is released. We obtain $E_{form}=-6.9$ eV. In Fig.\ 3(b) we address the electronic
band structure of the energetically favorable configuration 2/1b. We find that one spin
majority band and one spin minority band cross $E_F$, both contributed by the C $p_z$
orbitals. The semiconducting state of a monovacancy decorated by 2 O atoms is transferred
into a metallic state by B adsorption. 

For B adsorption at a monovacancy decorated by 3 O atoms, configuration 3/1a is characterized
by three ketone groups, establishing a BO$_3$ unit with an interatomic distance of
$d_{B-O}=1.38$ \AA. In configuration 3/1b the B atom is attached to one ketone group and
one C terminated ether group, which changes the original $sp^{2}$ hybridization into a
$sp^{3}$ hybridization, with interatomic distances of $d_{B-O}= 1.35$ \AA\ and
$d_{B-C}= 1.62$ \AA. Since in configuration 3/1c the B atom substitutes a C atom near the
vacancy, the structure does not change significantly. Configuration 3/1a is clearly
favorable with $E_{form}=-4.4$ eV, as compared to values of $-3.4$ eV and $-3.5$ eV for
configurations 3/1b and 3/1c. For configuration 3/1a we find a magnetic moment of
0.6 $\mu_B$. In addition, Fig. 3(c) shows that this configuration 3/1a is metallic with
two distinct C $p_z$ derived bands crossing $E_F$. The observed localized states
can be attributed to the under-coordination of the B atom, which clearly distinguishes this
case from configurations 3/1b and 3/1c.

In the case of 4 adsorbed O atoms, three configurations are found to be stable.
Configuration 4/1a has two ketone groups and one peroxide ($-$O$-$O$-$) group and
configuration 4/1b one ketone, one peroxide, and one ether group. The origin of the
stability is the fact that the BO$_3$ group itself is very stable \cite{new6}.
In the former case the B atom is adsorbed far away from to the monovacancy, while in the latter
it bonds with the 2 O atoms to form a BO$_3$ group. Configuration 4/1c is similar to
configuration 4/1a but with B substituting C at the boundary of the vacancy. Regarding 
the energetics, configuration 4/1b is strongly favorable with $E_{form}=-11.3$ eV, as
compared to values of $1.1$ eV and $1.6$ eV, since a peroxide group is present.
The electronic structure shown in Fig.\ 3(d) indicates that this configuration is
semiconducting with a band gap of 0.5 eV at the K point, and not magnetic. 

We next evaluate the effect of the B concentration on the electronic structure of
oxidized graphene by substituting 2 to 6 B atoms on C sites for a fixed O concentration
(two adsorbed O atoms), see Fig.\ 2. In configuration 2/2 two B atoms replace two 
C atoms at the boundary of the vacancy. The system develops a magnetic moment of 1 $\mu_B$
due to the presence of a dangling C bond. In configuration 2/3 the substitution of an
additional B atom does not change this situation. The fourth B atom in configuration 2/4
establishes a linear B$-$C=O bond, which is similar to the bond evolving in configuration
2/6. $E_{form}$ varies strongly between $-8.4$ eV and $-34.4$ eV due to the formation of
B$-$B dimers. Our results show that there is a strong tendency to B adsorption at oxidized
monovacancies.

We note the appearance of fractional magnetic moments in several cases: 0.1 $\mu_B$ in
configuration 2/1b and 0.2 $\mu_B$ in configurations 1/1b and 3/1a. For the remaining
systems the moment is either 0 or 1 $\mu_B$, see Table I, and therefore results from localized
dangling bonds. In all systems with fractional moments there are no contributions of
the adsorped O and/or B atoms to the magnetization. Spin polarization is carried only by
C atoms, where the atoms forming the boundary of the defect contribute less than those
further away. While clean monovacancies in graphene do not give rise to extended
$\pi$-band magnetism, the situation changes under O and B decoration.

\begin{figure*}[t]
\includegraphics[width=1\columnwidth]{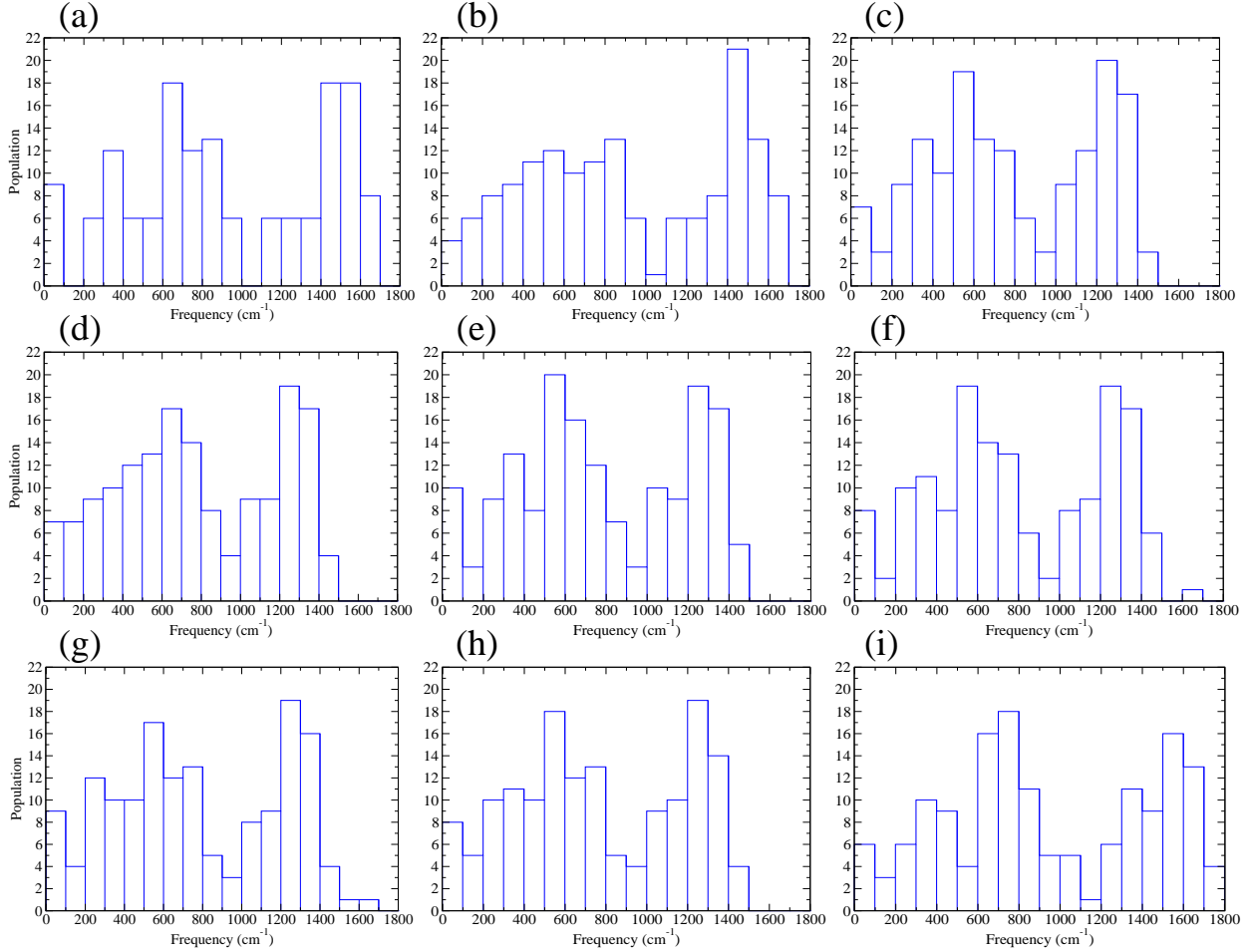}
\caption{\label{fig4} Phonon frequency at the $\Gamma$-point for (a) pristine graphene,
the structures (b) 1/1b, (c) 2/1b, (d) 3/1a, (e) 4/1b, (f) 2/2, 
(g) 2/4, (h) 2/6, and (i) 2 O atoms adsorbed at a monovacancy.}
\end{figure*}

The calculation of $\Gamma$-point phonons allows us to address the structural stability
after O and B doping. Phonon densities of states are shown as histograms for the most stable
structures (1/1b, 2/1b, 3/1a, 4/1b, 2/2, 2/4, 2/6) in Fig.\ 4 together with results for
pristine graphene and for 2 O atoms adsorbed at a monovacancy. Note that the reported
frequencies refers to the $\Gamma$-point of the Brillouine zone of our $5\times5$ supercell
and due to backfolding therefore also include frequencies of other points of the
standard graphene Brillouin zone. This phenomenon is well-known from carbon nanotubes
\cite{new7,new8,new9,new10,new11}. We find all phonon frequencies
to be positive and therefore conclude that all configurations addressed in Fig.\ 4 are
stable. Pristine graphene shows a two peak phonon spectrum, where the gross
shape is largely maintained under O and B adsorption. This finding is consistent with
recent experimental results of Raman spectroscopy of B doped graphene \cite{new5}.
However, there are characteristic
differences evident in the high frequency range beyond 1500 cm$^{-1}$. Except for
configuration 1/1b, the modes in this energy range are suppressed and the high frequency
peak shifts to the left. This fact cannot be a consequence of O adsorption, since an oxidized
monovacancy without adsorbed B atoms rather comes along with an enhancement of the high
energy modes, see Fig.\ 4(i). The softening of the G modes appears to be related
to out-of-plane shifts of O atoms and substantial local distortions in the graphene plane
induced by B adsorption.

{\bf 4. Conclusion}

We have performed first principles calculations to study the structural, electronic, and
magnetic properties of O and B decorated graphene. We have identified the energetically
favorable configurations for a variety of O and B concentrations and have demonstrated
that there exist magnetic solutions with and without dangling bonds. Since
vacancies in graphene are reactive and permit the incorporation of dopants, our calculations
demonstrate that B doping of oxidized vacancies is a successful approach to induce extended
$\pi$-band magnetism. By controlling the O and B concentrations, it is even possible to tune
the magnetic state. A study of the $\Gamma$-point phonons has been performed to understand
the structural stability of the decorated monovacancies.

{\bf Acknowledgments}

We thank KAUST research computing for providing the computational resources used for this investigation. 
M.\ Upadhyay Kahaly thanks SABIC for financial support. R.\ Faccio thanks the PEDECIBA, CSIC, and 
Agencia Nacional de Investigaci\'on e Innovaci\'on (ANII) Uruguayan organizations for financial support.

\end{document}